\documentclass{soups} 
\usepackage{booktabs} 

\usepackage{algorithm2e,amsmath}
\usepackage{balance}
\SetKwInOut{Input}{Input}
\SetKwInOut{Output}{Output\,}
\SetKwInOut{Data}{Data}
\SetKwProg{Tree}{Tree}{}{EndTree}
\usepackage{graphicx}
\usepackage{times}
\usepackage{balance}

\begin{document}

\conferenceinfo{
In: R. Verma, A. Das. (eds.): Proceedings of the 1st Anti-Phishing Shared Pilot at 4th ACM International Workshop on
Security and Privacy Analytics (IWSPA 2018)}{ Tempe, Arizona, USA, 21--03--2018, published at http://ceur-ws.org}

\title{A Deep Learning Model with Hierarchical LSTMs and Supervised Attention for Anti-Phishing}

\numberofauthors{3} 
\author{
% 1st. author
\alignauthor
Minh Nguyen\\
       \affaddr{Hanoi University of Science and Technology}\\
       \affaddr{Hanoi, Vietnam}\\
       \email{minh.nv142950@sis.hust.edu.vn}
% corresponding author
\alignauthor
Toan Nguyen\\
       \affaddr{New York University}\\
        \affaddr{Brooklyn, New York, USA}\\
       \email{toan.v.nguyen@nyu.edu}
 \and
% 3rd. author
\alignauthor
Thien Huu Nguyen\\
       \affaddr{University of Oregon}\\
        \affaddr{Eugene, Oregon, USA}\\
       \email{thien@cs.uoregon.edu}
}

\maketitle

\begin{abstract}
Anti-phishing aims to detect phishing content/documents in a pool of textual data. This is an important problem in cybersecurity that can help to guard users from fraudulent information. Natural language processing (NLP) offers a natural solution for this problem as it is capable of analyzing the textual content to perform intelligent recognition. In this work, we investigate state-of-the-art techniques for text categorization in NLP to address the problem of anti-phishing for emails (i.e, predicting if an email is phishing or not). These techniques are based on deep learning models that have attracted much attention from the community recently. In particular, we present a framework with hierarchical long short-term memory networks (H-LSTMs) and attention mechanisms to model the emails simultaneously at the word and the sentence level. Our expectation is to produce an effective model for anti-phishing and demonstrate the effectiveness of deep learning for problems in cybersecurity.
\end{abstract}

\keywords{Phishing, Deep Learning, NLP, H-LSTMs, Email Classification, Attentive LSTMs}

\section{Introduction}
Despite being one of the oldest tactics, email phishing remains the most common attack used by cybercriminals \cite{phishing2017} due to its effectiveness. Phishing attacks exploit users' inability to distinguish between legitimate information from fake ones sent to them~\cite{dhamija2006phishing, siadati2015verification, siadati2017x,siadati2017mind}. In an email phishing campaign, attackers send emails appearing to be from well-known enterprises or organizations directly to their victims or by spoofed emails~\cite{singer2005identification}. These emails try to lure victims to divulge their private information \cite{jagatic2007social,siadati2015verification,siadati2017mind} or to visit an impersonated site (i.e., a fake banking website), on which they will be asked for passwords, credit card numbers or other sensitive information. The recent hack of a high profile US politician (usually referred as ``John Podesta's hack'') is a famous example of this type of attack. It was all started by a spoofed email sent to the victim asking him to reset his Gmail password by clicking on a link in the email \cite{podesta}. The technique of email phishing may seem simple, yet the damage it makes is huge. In the US alone, the estimated cost of phishing emails to business is half a billion dollars per year \cite{phishingdamage}.

Numerous methods have been proposed to automatically detect phishing emails  \cite{bergholz2008improved,fette2007learning, abu2007comparison, gupta2017fighting}. Chandrasekaran et. al proposed to use structural properties of emails and Support Vector Machines (SVM) to classify phishing emails \cite{chandrasekaran2006phishing}. In  \cite{abu2007comparison}, Abu-Nimeh et. al evaluated six machine learning classifiers on a public phishing email dataset using proposed 43 features. Gupta et. al \cite{gupta2017fighting} presented a nice survey on recent state-of-the-art research on phishing detection. However, these methods mainly rely on feature engineering efforts to generate characteristics (features) to represent emails, over which machine learning methods can be applied to perform the task. Such feature engineering is often done manually and still requires much labor and domain expertise. This has hindered the portability of the systems to new domains and limited the performance of the current systems. 

In order to overcome this problem, our work focuses on deep learning techniques to solve the problem of phishing email detection. The major benefit of deep learning is its ability to automatically induce effective and task-specific representations from data that can be used as features to recognize phishing emails. As deep learning has been shown to achieve state-of-the-art performance for many natural language processing tasks, including text categorization  \cite{glorot2011domain,lai2015recurrent}, information extraction  \cite{nguyen2015relation,nguyen2015event,nguyen-grishman:2016:EMNLP2016}, machine translation \cite{bahdanau2014neural}, among others, we expect that it would also help to build effective systems for phishing email detection.

We present a new deep learning model to solve the problem of email phishing prediction using hierarchical long short-term memory networks (H-LSTMs) augmented with supervised attention technique. In the hierarchical LSTM model \cite{yang2016hierarchical}, emails are considered as hierarchical architectures with words in the lower level (the word level) and sentences in the upper level (the sentence level). LSTM models are first implemented in the word level whose results are passed to LSTM models in the sentence level to generate a representation vector for the entire email. The outputs of the LSTM models in the two levels are combined using the attention mechanism \cite{bahdanau2014neural} that assigns contribution weights to the words and sentences in the emails. A header network is also integrated to model the headers of the emails if they are available. In addition, we propose a novel technique to supervise the attention mechanism \cite{mi2016supervised,liu2016neural,liu2017exploiting} at the word level of the hierarchical LSTMs based on the appearance rank of the words in the vocabulary. Experiments on the datasets for phishing email detection in the First Security and Privacy Analytics Anti-Phishing Shared Task (IWSPA-AP 2018) \cite{ayman2018s} demonstrate the benefits of the proposed models, being ranked among the top positions among the participating systems of the shared task (in term of the performance on the unseen test data).

\section{Related Work}

Phishing email detection is a classic problem; however, research on this topic often has the same limitation: there is no official and big data set for it. Most previous works typically used a public set consists of legitimate or ``ham'' emails\footnote{https://spamassassin.apache.org/old/publiccorpus/} and another public set of phishing emails\footnote{https://monkey.org/~jose/phishing/} for their classification evaluation \cite{fette2007learning, bergholz2008improved, basnet2008detection, hamid2011hybrid, zhang2012phishing}. Other works used private but small data sets\cite{chandrasekaran2006phishing,abu2007comparison}. In addition, the ratio between phishing and legitimate emails in these data sets was typically balanced. This is not the case in the real-world scenario where the number of legitimate emails is much larger than that of phishing emails. Our current work relies on larger data sets with unbalanced distributions of phishing and legitimate emails collected for the the First Security and Privacy Analytics Anti-Phishing Shared Task (IWSPA-AP 2018) \cite{ayman2018s}.

Besides the limitation of small data sets, the previous work has extensively relied on feature engineering to manually find representative features for the problem. Apart from features extracted from emails, \cite{lakshmi2012efficient} also uses a blacklist of phishing webistes to get an additional feature for urls appearing in emails. Some neural network systems are also introduced to detect such blacklists \cite{mohammad2014predicting,martin2011framework}. This is undesirable because these engineered features need to be updated once new types of phishing emails with new content are presented. Our work differs from the previous work in this area in that we automate the feature engineering process using a deep learning model. This allows us to automatically learn effective features for phishing email detection from data. Deep learning has recently been employed for feature extraction with success on many natural language processing problems \cite{nguyen2015relation,nguyen2015event}.

\section{Proposed Model}

Phishing email detection is a binary classification problem that can be formalized as follow.

Let $e = \{b, s\}$ be an email in which $b$ and $s$ are the body content and header of the email respectively. Let $y$ the binary variable to indicate whether $e$ is a phishing email or not ($y=1$ if $e$ is a phishing email and $y=0$ otherwise). In order to predict the legitimacy of the email, our goal is to estimate the probability $P(y=1|e) = P(y=1|b,s)$. In the following, we will describe our methods to model the body $b$ and header $s$ with the body network and header network respectively to achieve this goal.

\subsection{Body Network with Hierarchical LSTMs}
\label{subsec:body}

For the body $b$, we view it as a sequence of sentences $b = (u_1,u_2,\ldots,u_L)$ where $u_i$ is the $i$-th sentence and $L$ is the number of sentences in the email body $b$. Each sentence $u_i$ is in turn a sequence of words/tokens $u_i = (v_{i,1}, v_{i,2}, \ldots, v_{i, K})$ with $v_{i,j}$ as the $j$-th token in $u_i$ and $K$ as the length of the sentence. Note that we set $L$ and $K$ to the fixed values by padding the sentences $u_i$ and the body $b$ with dummy symbols.

As there are two levels of information in $b$ (i.e, the word level with the words $v_{i,j}$ and the sentence level with the sentence $u_i$), we consider a hierarchical network that involves two layers of bidirectional long short-term memory networks (LSTMs) to model such information. In particular, the first layer consumes the words in the sentences via the embedding module, the bidirectional LSTM module and the attention module to obtain representation vectors for every sentence $u_i$ in $b$ (the word level layer). Afterward, the second network layer combines the representation vectors from the first layer with another bidirectional LSTM and attention module, leading to a representation vector for the whole body email $b$ (the sentence level layer). This body representation vector would then be used as features to estimate $P(y|b,s)$ and make the prediction for the initial email $e$.

\subsubsection{The Word Level Layer}
\label{sec:word-level}

\subsubsection*{Embedding}
In the word level layer, every word $v_{i,j}$ in each sentence $u_i$ in $b$ is first transformed into its embedding vector $w_{i,j}$. In this paper, $w_{i,j}$ is retrieved by taking the corresponding column vector in the word embedding matrix $W_e$ \cite{mikolov2013distributed} that has been pre-trained from a large corpus: $w_{i,j} = W_e[v_{i,j}]$ (each column in the matrix $W_e$ corresponds to a word in the vocabulary). As the result of this embedding step, every sentence $u_i = (v_{i,1}, v_{i,2}, \ldots, v_{i, K})$ in $b$ would be converted into a sequence of vectors $(w_{i,1}, w_{i,2}, \ldots, w_{i, K})$, constituting the inputs for the bidirectional LSTM model in the next step.

\subsubsection*{Bidirectional LSTMs for the word level} 
This module employs two LSTMs \cite{Hochreiter:97,graves2005framewise} that run over each input vector sequence $(w_{i,1}, w_{i,2}, \ldots, w_{i, K})$ via two different directions, i.e, forward (from $w_{i,1}$ to $w_{i,K}$) and backward (from $w_{i,K}$ to $w_{i,1}$). Along with their operations, the forward LSTM generates the forward hidden vector sequence $(\overrightarrow{h_{i,1}}, \overrightarrow{h_{i,2}}, \ldots, \overrightarrow{h_{i,K}})$ while the backward LSTM produce the backward hidden vector sequence $(\overleftarrow{h_{i,1}}, \overleftarrow{h_{i,2}}, \ldots, \overleftarrow{h_{i,K}})$. These two hidden vector sequences are then concatenated at each position, resulting in the new hidden vector sequence $(h_{i,1}, h_{i,2}, \ldots, h_{i,K})$ for the sentence $u_i$ in $b$ where $h_{i,j} = [\overrightarrow{h_{i,j}}, \overleftarrow{h_{i,j}}]$. The notable characteristics of the hidden vector $h_{i,j}$ is that it encodes the context information over the whole sentence $u_i$ due to the recurrent property of the forward and backward LSTMs although a greater focus is put at the current word $v_{i,j}$.

\begin{figure}[h]
\includegraphics[scale=0.175]{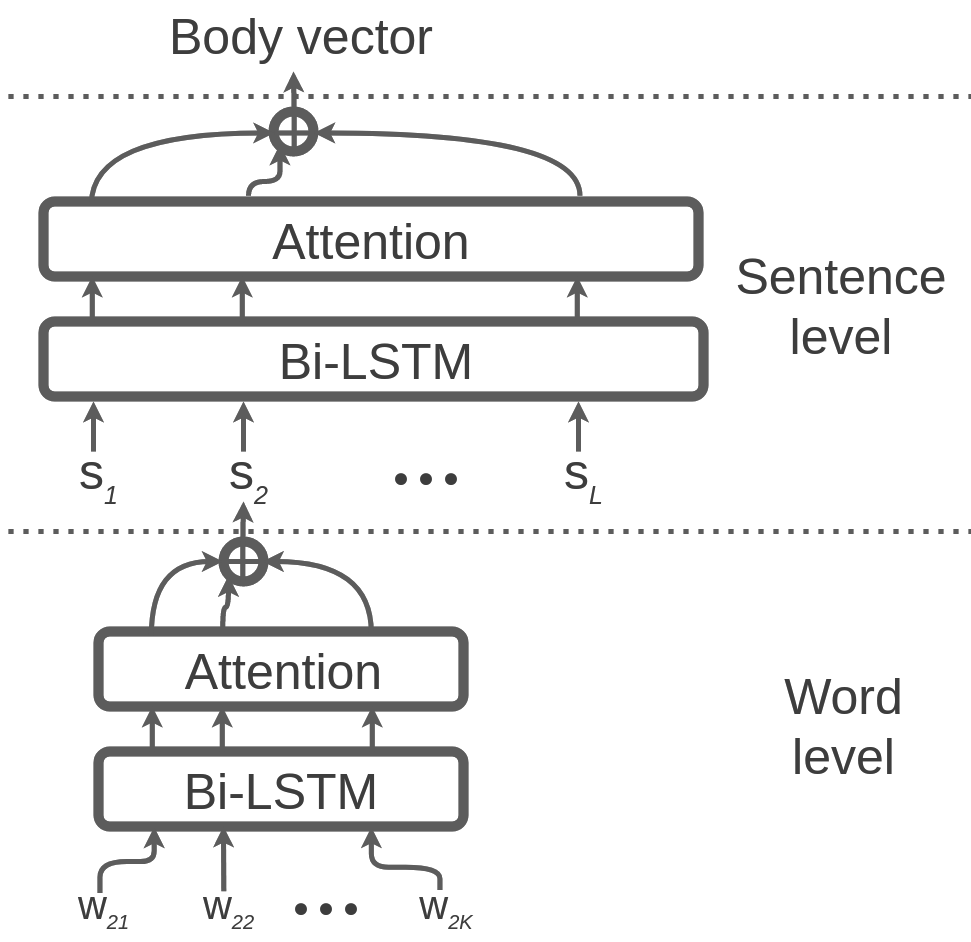}
\centering
\caption{Hierarchical LSTMs.}
\label{fig:analysis-1}
\end{figure}

\subsubsection*{Attention}

In this module, the vectors in the hidden vector sequence $(h_{i,1}, h_{i,j,2}, \ldots, h_{i,K})$ are combined to generate a single representation vector for the initial sentence $u_i$. The attention mechanism \cite{yang2016hierarchical} seeks to do this by computing a weighted  sum of the vectors in the sequence. Each hidden vector $h_{i,j}$ would be assigned to a weight $\alpha_{i,j}$ to estimate its importance/contribution in the representation vector for $u_i$ for the phishing prediction of the email $e$. In this work, the weight $\alpha_{i,j}$ for $h_{i,j}$ is computed by:

\begin{equation}
\label{eq:learned_weight}
\alpha_{i,j}=\frac{\exp(a_{i,j}^\top w_{a})}{\sum_{j'}{\exp(a_{i,j'}^\top w_{a})}}
\end{equation}
in which
\begin{equation}
\label{eq:att_rep}
a_{i,j}=\tanh(W_{att} h_{i,j} + b_{att})
\end{equation}
Here, $W_{att}$, $b_{att}$ and $w_{a}$ are the model parameters that would be learnt during the training process. Consequently, the representation vector $\hat{u}_i$ for the sentence $u_i$ in $b$ would be:

\begin{equation}
\label{eq:sent_rep}
\hat{u}_i=\sum_j\alpha_{i,j}h_{i,j}
\end{equation}

After the word level layer completes its operation on every sentence of $b=(u_1, u_2, \ldots, u_L)$, we obtain a corresponding sequence of sentence representation vectors $(\hat{u}_1, \hat{u}_2, \ldots, \hat{u}_L)$. This vector sequence would be combined in the next sentence level layer to generate a single vector to represent $b$ for phishing prediction.

\subsubsection{The Sentence Level Layer}

The sentence level layer processes the vector sequence $(\hat{u}_1, \hat{u}_2, \ldots, \hat{u}_L)$ in the same way that the word level layer has employed for the vector sequence  $(w_{i,1}, w_{i,2}, \ldots, w_{i, K})$ for each sentence $u_i$. Specifically, $(\hat{u}_1, \hat{u}_2, \ldots, \hat{u}_L)$ is also first fed into a bidirectional LSTM module (i.e, with a forward and backward LSTM) whose results are concatenated at each position to produce the corresponding hidden vector sequence $(\hat{h}_1, \hat{h}_2, \ldots, \hat{h}_L)$. In the next step with the attention module, the vectors in $(\hat{h}_1, \hat{h}_2, \ldots, \hat{h}_L)$ are weighted and summed to finally generate the representation vector $r^b$ for the email body $b$ of $e$. Assuming the attention weights for $(\hat{h}_1, \hat{h}_2, \ldots, \hat{h}_L)$ are $(\beta_1, \beta_2, \ldots, \beta_L)$ respectively. the body vector $r_b$ is then computed by:

\begin{equation}
\label{eq:body_rep}
r^b=\sum_i\beta_i\hat{h}_i
\end{equation}

Note that the model parameters of the bidirectional LSTM modules (and the attention modules) in the word level layer and the sentence level layer are separate and they are both learnt in a single training process. Figure \ref{fig:analysis-1} shows the overview of the body network with hierarchical LSTMs and attention.

Once the body vector $r^b$ has been computed, we can use it as features to estimate the phishing probability via:

\begin{equation}
\label{eq:prob}
P(y=1|b,s) = \sigma(W_{out}r^b + b_{out})
\end{equation}

where $W_{out}$ and $b_{out}$ are the model parameters and $\sigma$ is the logistic function.

\subsection{Header Network}
\label{subsubsec:h-LSTMs-header}

\begin{figure}[h]
\includegraphics[scale=0.175]{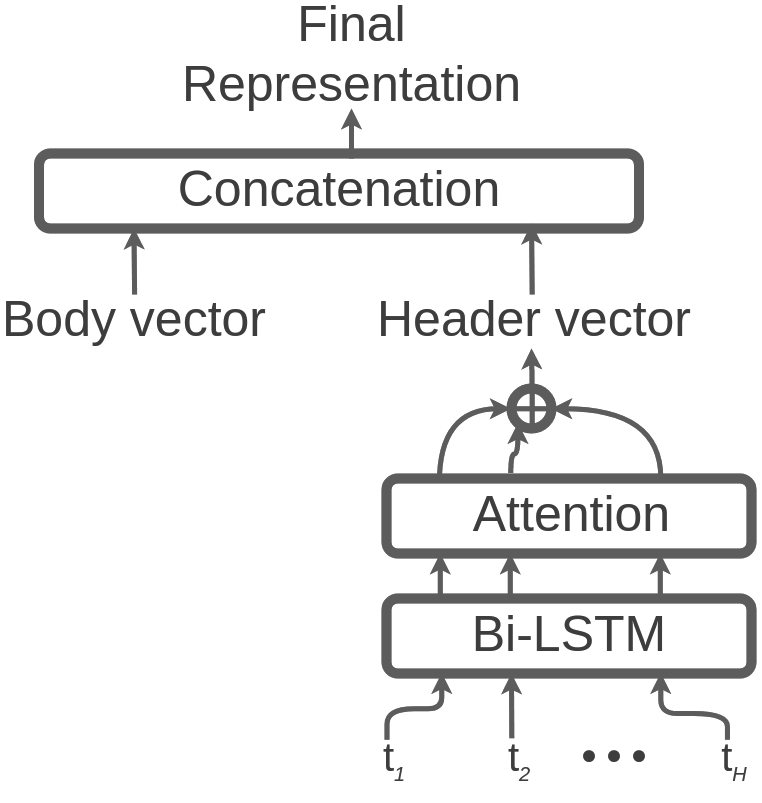}
\centering
\caption{Hierarchical LSTMs with header network.}
\label{fig:analysis-2}
\end{figure}

The probability estimation in Equation \ref{eq:prob} does not consider the headers of the emails. For the email datasets with headers available, we can model the headers with a separate network and use the resulting representation as additional features to estimate the phishing probability. In this work, we consider the header $s$ of the initial email $e$ as a sequence of words/tokens: $(x_i, x_2, \ldots, x_H)$ where $x_i$ is the $i$-th word in the header and $H$ is the length of the header. In order to compute the representation vector $r^s$ for $s$, we also employ the same network architecture as the word level layer in the body network using separate modules for embedding module, bidirectional LSTM, and attention (i.e, Section \ref{sec:word-level}). An overview of this header network is presented in Figure \ref{fig:analysis-2}.

Once the header representation vector $r^s$ is generated, we concatenate it with the body representation vector $r^b$ obtained from the body network, leading to the final representation vector $r = [r^b, r^s]$ to compute the probability $P(y=1|b,s) = \sigma(W_{sub}r + b_{sub})$ ($W_{sub}$ and $b_{sub}$ are model parameters).

In order to train the models in this work, we minimize the negative log-likelihood of the models on a training dataset in which the negative log-likelihood for the email $e$ is computed by:

\begin{equation}
\label{eq:original_loss}
L_c = - \log(P(y=1|e))
\end{equation}

The model we have described so far is called H-LSTMs for convenience.

\subsection{Supervised Attention}
\label{subsubsec:supervised}

The attention mechanism in the body and header networks is expected to assign high weights for the informative words/sentences and downgrade the irrelevant words/sentences for phishing detection in the emails. However, this ideal operation can only be achieved when an enormous training dataset is provided to train the models. In our case of phishing email detection, the size of the training dataset is not large enough and we might not be able to exploit the full advantages of the attention. In this work, we seek for useful heuristics for the problem and inject them into the models to facilitate the operation of the attention mechanism. In particular, we would first heuristically decide a score for every word in the sentences so that the words with higher scores are considered as being more important for phishing detection than those with lower scores. Afterward, the models would be encouraged to produce attention weights for words that are close to their heuristic importance scores. The expectation is that this mechanism would help to introduce our intuition into the attention weights to compensate for the small scale of the training dataset, potentially leading to a better performance of the models. Assuming the importance scores for the words in the sentence $(v_{i,1}, v_{i,2}, \ldots, v_{i,K})$ be $(g_{i,1}, g_{i,2}, \ldots, g_{i,K})$ respectively, we force the attention weights $(\alpha_{i,1}, \alpha_{i,2}, \ldots, \alpha_{i,K})$ (Equation \ref{eq:learned_weight}) to be close to the importance scores by penalizing the models that render large square difference between the attention weights and the importance scores. This amounts to adding the square difference into the objective function in Equation \ref{eq:original_loss}:

\begin{equation}
\label{eq:supervisedLoss}
L_e = L_c + \lambda \sum_{i,j}(g_{i,j} - \alpha_{i,j})^2
\end{equation}

where $\lambda$ is a trade-off constant.

\subsubsection*{Importance Score Computation}

In order to compute the importance scores, our intuition is that a word is important for phishing detection if it appears frequently in phishing emails and less frequently in legitimate emails. The fact that an important word does not appear in many legitimate emails helps to eliminate the common words that are used in most documents. Consequently, the frequent words that are specific to the phishing emails would receive higher importance scores in our method. Note that our method to find the important words for phishing emails is different from the prior work that has only considered the most frequent words in the phishing emails and ignored their appearance in the legitimate emails.

We compute the importance scores as follow. For every word $v$ in the vocabulary, we count the number of the phishing and legitimate emails in a training dataset that contain the word. We call the results as the phishing email frequency and the legitimate email frequency respectively for $v$. In the next step, we sort the words in the vocabulary based on its phishing and legitimate email frequencies in the descending order. After that, a word $v$ would have a phishing rank ($phishingRank(v)$) and a legitimate rank ($legitimateRank(v)$) in the sorted word sequences based on the phishing and legitimate frequencies (the higher the rank is, the less the frequency is). Given these ranks, the unnormalized importance score for $v$ is computed by:\footnote{The actual important scores of the words we use in Equation \ref{eq:supervisedLoss} are normalized for each sentence.}

\begin{equation}
\label{eq:posWeight}
score[v] = \frac{legitimateRank[v]}{phishingRank[v]}
\end{equation}

The rationale for this formula is that a word would have a high importance score for phishing prediction if its legitimate rank is high and its phishing rank is low. Note that we use the ranks of the words instead of the frequencies because the frequencies are affected by the size of the training dataset, potentially making the scores unstable. The ranks are less affected by the dataset size and provide a more stable measure. Table \ref{tab:top-weight} demonstrates the top 20 words with the highest unnormalized importance scores in our vocabulary.

\begin{table}[h]
\centering
\caption{Top 20 words with the highest scores.}
\label{tab:top-weight}
\begin{tabular}{|l|l|}
\hline
account  & 21.45 \\ \hline
your     & 15.00 \\ \hline
click    & 14.11 \\ \hline
mailbox  & 9.59  \\ \hline
cornell  & 9.58  \\ \hline
link     & 9.37  \\ \hline
verify   & 8.83  \\ \hline
customer & 8.63  \\ \hline
access   & 8.50  \\ \hline
reserved & 8.03  \\ \hline
dear     & 7.85  \\ \hline
log      & 7.70  \\ \hline
accounts & 7.61  \\ \hline
paypal   & 7.52  \\ \hline
complete & 7.37  \\ \hline
service  & 7.15  \\ \hline
protect  & 6.95  \\ \hline
secure   & 6.94  \\ \hline
mail     & 6.70  \\ \hline
clicking & 6.63  \\ \hline
\end{tabular}
\end{table}

The H-LSTMs model augmented with the supervised attention mechanism above is called H-LSTM+supervised in the experiments.

\subsubsection{Training}

We train the models in this work with stochastic gradient descent, shuffled mini-batches, Adam update rules \cite{Kingma:14}. The gradients are computed via back-propagation while dropout is used for regularization \cite{Srivastava:14}. We also implement gradient clipping to rescale the Frobenius norms of the non-embedding weights if they exceed a predefined threshold.

\section{Evaluation}
\subsection{Datasets and Preprocessing}

The models in this work are developed to participate in the First Security and Privacy Analytics Anti-Phishing Shared Task (IWSPA-AP 2018) \cite{ayman2018s}. The organizers provide two datasets to train the models for email phishing recognition. The first dataset involves emails that only have the body part (called \textit{data-no-header}) while the second dataset contains emails with both bodies and headers (called \textit{data-full-header}. These two datasets translate into two shared tasks to be solved by the participants. The statistics of the training data for these two datasets are shown in Table \ref{tab:datasets}.

\begin{table}[h]
\centering

\begin{tabular}{lcc}
\hline
\textbf{Datasets} & \textbf{\#legit} & \textbf{\#phish} \\ \hline
\textit{data-no-header}   & 5092                  & 629                 \\ \hline
\textit{data-full-header} & 4082                  & 503                 \\ \hline
\end{tabular}
\caption{Statistics of the \textit{data-no-header} and \textit{data-full-header} datasets. \textbf{\#legit} and \textbf{\#phish} are the numbers of legitimate and phishing emails respectively.}
\label{tab:datasets}
\end{table}

The raw test data (i.e, without labels) for these datasets are released to the participants at a specified time. The participants would have one week to run their systems on such raw test data and submit the results to the organizers for evaluation.

%As there is no development data provided, we need to take parts of the training data to develop the models.

\begin{figure}[h]
\includegraphics[scale=0.175]{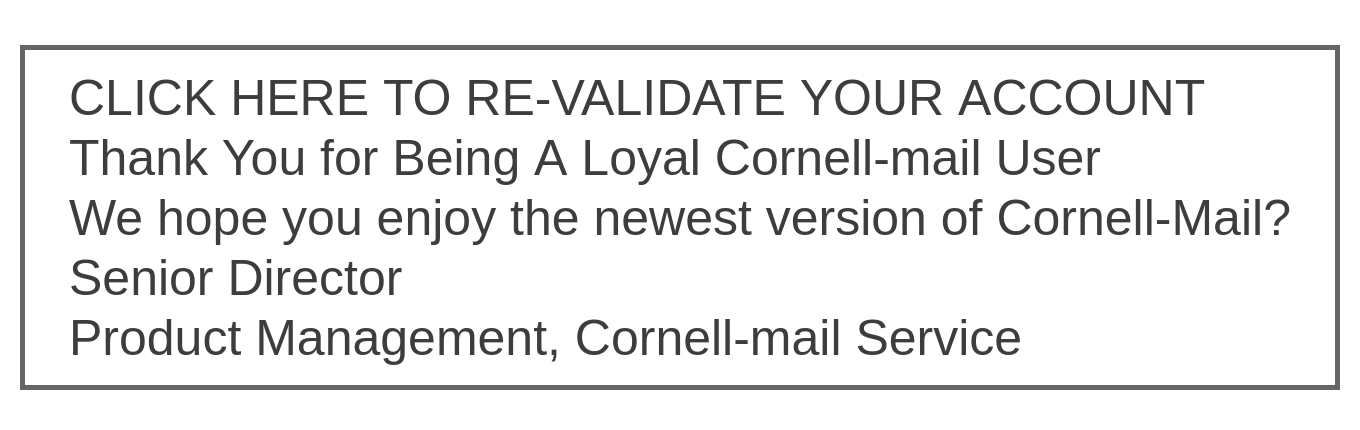}
\centering
\caption{A case in which splitting body into sentences cannot be done as usual. (Phishing email: 28.txt in data-no-header).}
\label{fig:sentence-splitting}
\end{figure}

Regarding the preprocessing procedure for the datasets, we notice that a large part of the text in the email bodies is quite unstructured. The sentences are often short and/or not clearly separated by the ending-sentence symbols (i.e, $\{$. ! ?$\}$). In order to split the bodies of the emails into sentences for our models, we develope an in-house sentence splitter specially designed for the datasets. In particular, we determine the beginning of a sentence by considering if the first word of a new line is capitalized or not, or if a capitalized word is immediately followed by an ending-sentence symbol. The sentences whose lengths (numbers of words) are less than 3 are combined to create a longer sentence. This reduces the number of sentences significantly and expands the context for the words in the sentences as they are processed by the models. Figure \ref{fig:sentence-splitting} shows a phishing email from the datasets.

\subsection{Baselines}

In order to see how well the proposed deep learning models (i.e, H-LSTMs and H-LSTMs+supervised) perform with respect to the traditional methods for email phishing detection, we compare the proposed models with a baseline model based on Support Vector Machines (SVM) \cite{chandrasekaran2006phishing}. We use the \texttt{tf-idf} scores of the words in the vocabulary as the features for this baseline \cite{chandrasekaran2006phishing}. Note that since the email addresses and urls in the provided datasets have been mostly hidden to protect personal information, we cannot use them as features in our SVM baselines as do the previous systems.

We employ the implementation of linear and nonlinear SVM from the \texttt{sklearn} library \cite{scikit-learn} for which the \texttt{tf-idf} representations of the emails are obtained via the \texttt{gensim} toolkit \cite{rehurek_lrec}.

\subsection{Hyper-parameter Selection}

As the size of the provided datasets is small and no development data is included, we use a 5-fold stratified cross validation on the training data of the provided datasets to search for the best hyper-parameters for the models. The hyper-parameters we found are as follows.

The size of word embedding vector is 300 while the cell sizes are set to 60 for all the LSTMs in the body and header networks. The size of attention vectors at the attention modules for the body and header networks are also set to 60. The $\lambda$ coefficient for supervised attention is set to 0.1, the threshold for gradient clipping is 0.3 and the drop rate for drop-out is 0.5. For the Adam update rule, we use the learning rate of 0.0025. Finally, we set $C=10.0$ for the linear SVM baseline. The nonlinear version of SVM we use is C-SVC with radial basis function kernel and $(C, \gamma)=(50.0, 0.1)$.

\subsection{Results}

In the experiments below, we employ the precision, recall and F1-score to evaluate the performance of the models for detecting phishing emails. In addition, the proposed models H-LSTMs and H-LSTMs+supervised only utilize the header network in the evaluation on \textit{data-full-header}.

\subsubsection*{Data without header}
In the first experiment, we compare the proposed models with the SVM baselines in two different settings when the email headers are not considered (the first shared task). In particular, in the first setting, we use \textit{data-no-header} as the training data and perform a 5-fold stratified cross-validation to evaluate the models. In the second setting, \textit{data-no-header} is also utilized as the training data, but the bodies extracted from \textit{data-full-header} (along with the corresponding labels) are employed as the test data. The results of the first setting are shown in Table \ref{tab:cross-no-headers} while the results of the second setting are presented in Table \ref{tab:test-full-headers}.

\begin{table}[h]
\centering
\begin{tabular}{lccc}
\hline
\textbf{Models}  & \textbf{Precision} & \textbf{Recall}  & \textbf{F1}  \\ \hline
H-LSTMs+supervised & 0.9784             & 0.9466 & 0.9621                \\ \hline
H-LSTMs          & 0.9638             & 0.9448   & 0.9542               \\ \hline
Linear SVM+tfidf  & 0.9824 & 0.8856 & 0.9313 \\ \hline
Linear SVM+embedding  & 0.9529 & 0.9206 & 0.9364 \\ \hline
Linear SVM+tfidf+embedding  & 0.9837 & 0.9253 & 0.9536 \\ \hline
Kernel SVM+tfidf & 0.9684 & 0.8730 & 0.9180 \\ \hline
Kernel SVM+embedding & 0.9408 & 0.9141 & 0.9273 \\ \hline
Kernel SVM+tfidf+embedding & 0.9714 & 0.9174 & 0.9436 \\ \hline
\end{tabular}
\caption{Performance comparison between the proposed models H-LSTMs and H-LSTMs+supervised with the baseline models Linear and Kernel SVM.}
\label{tab:cross-no-headers}
\end{table}

\begin{table}[h]
\centering
\caption{Performance of all models on the test data (data-full-headers).}
\label{tab:test-full-headers}
\begin{tabular}{lccc}
\hline
\textbf{Models} & \textbf{Precision} & \textbf{Recall}  & \textbf{F1}   \\ \hline
H-LSTMs+supervised   & 0.8892             & 0.7395 & 0.8075              \\ \hline
H-LSTMs        & 0.8934             & 0.7054      & 0.7883              \\ \hline
Linear SVM+tfidf & 0.8864 & 0.6978 & 0.7809 \\ \hline
Linear SVM+embedding  & 0.8112 & 0.6918 & 0.7468 \\ \hline
Linear SVM+tfidf+embedding  & 0.8695 & 0.7018 & 0.7767 \\ \hline
Kernel SVM+tfidf & 0.8698 & 0.7038 & 0.7780 \\ \hline
Kernel SVM+embedding & 0.8216 & 0.6501 & 0.7259 \\ \hline
Kernel SVM+tfidf+embedding & 0.8564 & 0.6937 & 0.7665 \\ \hline
\end{tabular}
\end{table}

As we can see from the tables, the two versions of hierarchical LSTMs (i.e, H-LSTMs and H-LSTMs+supervised) outperform the baseline models in both experiment settings. The performance improvement is significant with large margins (up to 3\% improvement on the absolute F1 score in the first experiment setting). The main gain is due to the recall, demonstrating the generalization advantages of the proposed deep learning models over the traditional methods for phishing detection with SVM. Comparing H-LSTMs+supervised and H-LSTMs, we see that H-LSTMs+supervised is consistently better than H-LSTMs with significant improvement in the second setting. This shows the benefits of supervised attention for hierarchical LSTM models for email phishing detection. Finally, we see that the performance in the first setting is in general much better than those in the second setting. We attribute this to the fact that text data in \textit{data-no-header} and \textit{data-full-header} is quite different, leading to the mismatch between data distributions of the training data and test data in the second experiment setting.

In the final submission for the first shared task (i.e, without email headers), we combine the training data from \textit{data-no-header} with the extracted bodies (along with the corresponding labels) from the training data of \textit{data-full-header} to generate a new training set. As H-LSTM+supervised is the best model in this development experiment, we train it on the new training set and use the trained model to make predictions for the actual test set of the first shared task.

\subsubsection*{Data with full header} 

In this experiment, we aim to evaluate if the header network can help to improve the performance of H-LSTMs. We take the training dataset from \textit{data-full-header} to perform a 5-fold cross-validation evaluation. The performance of H-LSTMs when the header network is included or excluded is shown in Table \ref{tab:cross-full-headers}.

\begin{table}[h]
\centering
\caption{Cross-validation performance of H-LSTMs with using headers compared to the original version.}
\label{tab:cross-full-headers}
\begin{tabular}{lccc}
\hline
\textbf{Models}   & \textbf{Precision} & \textbf{Recall}    & \textbf{F1}\\ \hline
H-LSTMs (only body)   & 0.9732             & 0.9534   & 0.9631            \\ \hline
H-LSTMs + headers  & 0.9816             & 0.9596 & 0.9705                \\ \hline
\end{tabular}
\end{table}

From the table, we see that the header network is also helpful for H-LSTMs as it helps to improve the performance of H-LSTMs for the dataset with email headers (an 0.7\% improvement on the F1 score).

In the final submission for the second shared task (i.e, with email headers), we simply train our best model in this setting (i.e, H-LSTM+supervised) on the training dataset of \textit{data-full-header}.

The time for the training and test process of the proposed models is shown in the Table \ref{tab:time}. Note that the training time of \textit{H-LSTMs+headers+supervised} is longer than that of \textit{H-LSTMs+supervised} since the first model's training data includes both the original training data of the first task and the extracted bodies from the training data of the second task. The test data of the first task with \textit{H-LSTMs+supervised} is also larger than that of the second task with\\ \textit{H-LSTMs+headers+supervised}.

\begin{table}[h]
\centering
\caption{Training and test times of our models.}
\label{tab:time}
\begin{tabular}{lll}
\hline
\textbf{Models}            & \textbf{Training time} & \textbf{Test time} \\ \hline
H-LSTMs+supervised         & \multicolumn{1}{c}{3.7 hours}              & \multicolumn{1}{c}{4 minutes}          \\ \hline
H-LSTMs+headers+supervised & \multicolumn{1}{c}{1.5 hours}              & \multicolumn{1}{c}{1 minute}          \\ \hline
\end{tabular}
\end{table}
\subsubsection*{Comparision with the participating systems on the actual test sets} 

Tables \ref{tab:test-data-no-header} and \ref{tab:test-data-full-header} show the best performance on the actual test data of all the teams that participate in the shared tasks. Table \ref{tab:test-data-no-header} reports the performance for the first shared task (i.e, without email headers) while Table \ref{tab:test-data-full-header} presents the performance for the second shared task (i.e, with email headers). These performance is measured and released by the organizers. The performance of the systems we submitted is shown in the rows with our team name (i.e, \textit{TripleN}).
\begin{table}[h]
\centering
\caption{The best performance of all the participating teams in the first shared task with no email headers.}
\begin{tabular}{lccc}
\hline
\multicolumn{1}{c}{\textbf{Teams}} & \textbf{Precision} & \textbf{Recall} & \textbf{F1}  \\ \hline
\textbf{TripleN (our team)}            & \textbf{0.981}     & \textbf{0.978}   & \textbf{0.979}   \\ \hline
Security-CEN@Amrita        & 0.962              & 0.989        & 0.975                  \\ \hline
Amrita-NLP                      & 0.972              & 0.974     & 0.973                   \\ \hline
CEN-DeepSpam                & 0.951              & 0.964        & 0.958                    \\ \hline
CENSec@Amrita               & 0.914              & 0.998       & 0.954                     \\ \hline
CEN-SecureNLP              & 0.890              & 1.0         & 0.942                      \\ \hline
CEN-AISecurity            & 0.936              & 0.910          & 0.923                    \\ \hline
Crypt Coyotes                & 0.936              & 0.910      & 0.923                     \\ \hline
\end{tabular}
\label{tab:test-data-no-header}
\end{table}

\begin{table}[h]
\centering
\caption{The best performance of all the participating teams in the second shared task with email headers.}
\begin{tabular}{lccc}
\hline
\multicolumn{1}{c}{\textbf{Teams}} & \textbf{Precision} & \textbf{Recall} & \textbf{F1}     \\ \hline
Amrita-NLP             & 0.998     & 0.994   & 0.996   \\ \hline
\textbf{TripleN (our team)}                & \textbf{0.990}              & \textbf{0.992}          & \textbf{0.991}                    \\ \hline
CEN-DeepSpam                 & 1.000              & 0.978       & 0.989                    \\ \hline
Security-CEN@Amrita             & 0.998              & 0.976     & 0.987                   \\ \hline
CENSec@Amrita               & 0.882              & 1.000        & 0.937                    \\ \hline
CEN-AISecurity                    & 0.957              & 0.900     & 0.928                 \\ \hline
CEN-SecureNLP                 & 0.880              & 0.971        & 0.924                  \\ \hline
Crypt Coyotes                  & 0.960              & 0.863          & 0.909               \\ \hline
\end{tabular}
\label{tab:test-data-full-header}
\end{table}

As we can see from the tables, our systems achieve the best performance for the first shared task and the second best performance for the second shared task. These results are very promising and demonstrate the advantages of the proposed methods in particular and deep learning in general for the problem of email phishing recognition.

\section{Conclusions}

We present a novel deep learning model to detect phishing emails. Our model employs hierarchical attentive LSTMs to model the email bodies at both the word level and the sentence level. A header network with attentive LSTMs is also incorporated to model the headers of the emails. In the model, we propose a novel supervised attention technique to improve the performance using the email frequency ranking of the words in the vocabulary. Several experiments are conducted to demonstrate the benefits of the proposed deep learning model.

\balance
\bibliographystyle{abbrv}

\end{document}